\documentclass[10pt,preprint2]{aastex}
\usepackage{wasysym}

\def\planet{KOI-13.01}
\def\star{KOI-13}

\def\asc{\lambda}
\def\obliq{\psi}
\def\spinorbit{\varphi}

\shorttitle{Orbit Inclination of \planet}
\shortauthors{Jason W. Barnes}

\begin{document}

\title{Measurement of the Spin-Orbit Misalignment of \planet~from its 
Gravity-Darkened \emph{Kepler} Transit Lightcurve}

\author{Jason W. Barnes} 
\affil{Department of Physics} 
\affil{University of Idaho}
\affil{Moscow, ID 83844-0903, USA} 
\affil{ResearcherID:  B-1284-2009} 
\email{jwbarnes@uidaho.edu}

\author{Ethan Linscott} 
\affil{Department of Physics} 
\affil{Oklahoma Baptist University}
\affil{Shawnee, OK 74804, USA}

\author{Avi Shporer} 
\affil{Department of Physics} 
\affil{University of California, Santa Barbara}
\affil{Santa Barbara, CA 93106, USA}
\affil{}
\affil{Las Cumbres Observatory Global Telescope Network}
\affil{Santa Barbara, CA 83117, USA}

\newpage

\begin{abstract} 

We model the asymmetry of the \planet~transit lightcurve assuming a gravity-darkened
rapidly-rotating host star in order to constrain the system's spin-orbit alignment and
transit parameters.  We find that our model can reproduce the \emph{Kepler}
lightcurve for \planet~with a sky-projected alignment of $\asc=23^\circ\pm4^\circ$ and
with the star's north pole tilted away from the observer by $48^\circ\pm4^\circ$
(assuming $M_*=2.05~\mathrm{M_\odot}$).  With both these determinations, we
calculate that the net misalignment between this planet's orbit normal and its
star's rotational pole is $56^\circ\pm4^\circ$.  Degeneracies in our geometric
interpretation also allow a retrograde spin-orbit angle of
$124^\circ\pm4^\circ$.  This is the first spin-orbit measurement to come from
gravity darkening, and is one of only a few measurements of the full (not just the
sky-projected) spin-orbit misalignment of an extrasolar planet.  We also measure
accurate transit parameters incorporating stellar oblateness and gravity darkening: 
$R_*=1.756\pm0.014~\mathrm{R_\odot}$, $R_\mathrm{p}=1.445\pm0.016~\mathrm{R_{Jup}}$,
and $i=85.9^\circ\pm0.4^\circ$.  The new lower planetary radius falls within the
planetary mass regime for plausible interior models for the transiting body. 
A simple initial calculation shows that \planet's circular orbit is 
apparently inconsistent with 
the Kozai mechanism having driven its
spin-orbit misalignment; planet-planet scattering and stellar spin migration remain
viable mechanisms.  Future \emph{Kepler} data will improve the precision of the
\planet~transit lightcurve, allowing more precise determination of transit
parameters and the opportunity to use the Photometric Rossiter-McLaughlin effect to
resolve the prograde/retrograde orbit determination degeneracy.

\end{abstract}

 
\section{INTRODUCTION}

\begin{figure}[h]
\epsscale{1.}
\plotone{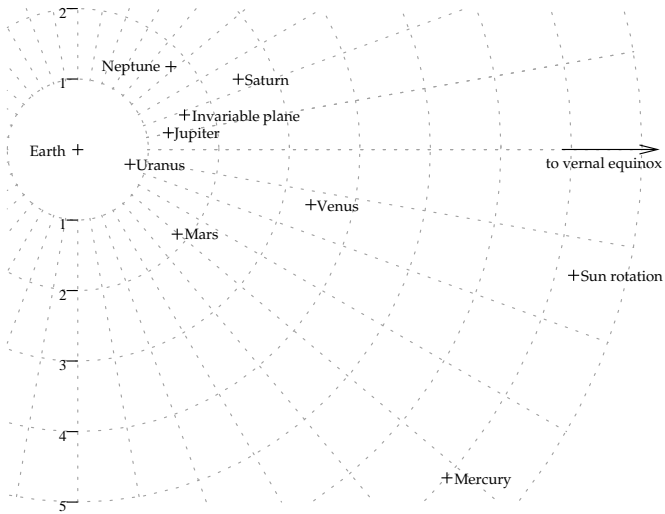}
\caption{\footnotesize Relative positions of the angular momentum vectors for the orbits of
solar system planets, along with that for the Sun's rotation and the solar system
invariable plane (plane of the net angular momentum), are plotted in J2000 ecliptic
coordinates.  Planetary orbit locations calculated from orbital elements from
\citet{SolarSystemDynamics}, Sun's rotation pole calculated from equation given in \citet{1992exaa.book.....S}, and the
invariable plane coordinates are from \citet{AstrophysicalQuantities}.
\label{figure:Lvectors}}
\end{figure}

The Sun's planets all orbit in planes that vary in orbital inclination from one
another by just $\sim1^\circ-7^\circ$ -- a situation that supports the nebular
hypothesis for the solar system's origin \citep[see,
\emph{e.g.}][]{1993ARA&A..31..129L}.  The orbital planes all differ by between
$\sim3^\circ-7^\circ$ from the Sun's equatorial plane, and by
$\sim0.3^\circ-6.3^\circ$ from the invariable plane (the plane normal to the net
solar system angular momentum vector).  Although angular distances are typically
provided for these differences, we think that they are better shown in two
dimensions.  Figure \ref{figure:Lvectors} shows the location of the orbital angular
momentum vectors for various solar system objects in ecliptic coordinates.

Given its utility in constraining the origin of the solar system, measurement of
the mutual inclination between extrasolar planets' orbits and/or the inclination
of extrasolar planets' orbit normal relative to their stars' spin axis --
`spin-orbit alignment', $\spinorbit$ -- can shed light on the origin of those
extrasolar systems as well.  Evidence so far shows that while the solar
system is not unusual, a wide variety of planetary system types exists.

Some systems have planets that are even more nearly coaligned with each other than
those in the solar system.  In particular, the Kepler-11 six-planet system has
typical mean mutual inclinations of just $1^\circ-2^\circ$
\citep{2011Natur.470...53L}.  As evidenced by Figure \ref{figure:Lvectors},
$1^\circ-2^\circ$ mutual inclination is significantly smaller than the mutual
inclination of the solar system's terrestrial planets.  Interestingly, however, the
solar system giant planets have mutual inclinations of $0.7^\circ-2.0^\circ$,
comparable to those in the Kepler-11 system.  We think that this similarity in
mutual inclinations between the solar system giants and the Kepler-11 planets may
indicate similar formation and evolution, but distinctly different from that of the
solar system terrestrial planets.  Although coplanarity of a system's planets and a
planet's spin-orbit alignment are distinct measurements, they share the property
that the proposed mechanisms for misaligning one will likely misalign the other as
well.

Radial velocity measurements of planet-hosting stars during transits can indicate
spin-orbit alignment via the Rossiter-McLaughlin effect.  Rossiter-McLaughlin
observations have revealed that a majority of close-in extrasolar planets appear to
be spin-orbit aligned \citep[20 of 31 measurements of $\asc$ tabulated by][page 129,
are within $2\sigma$ of zero]{2011exha.book.....P}.  For example,
\citet{2006ApJ...653L..69W} showed that HD189733 has a spin-orbit misalignment of
$1.4^\circ\pm1.1^\circ$, which \citet{2009A&A...506..377T} later refined to just
$0.85^\circ\pm0.30^\circ$.  HAT-P-1's measured inclination of $3.7^\circ \pm
2.1^\circ$ is consistent with spin-orbit alignment \citep{2008ApJ...686..649J}.  We
note that, based on Figure \ref{figure:Lvectors}, each of these planets is more
closely spin-orbit aligned than is Jupiter, which has an inclination of $6.1^\circ$
with respect to the Sun's equatorial plane.  Many more planets have been determined
to be spin-orbit aligned to lower precision  \citep[\emph{e.g.}][to name a few recent
measurements]{2010MNRAS.405.1867S,2010A&A...523A..52M,2011ApJ...730L..31H,2011MNRAS.414.3023S}.

On the other hand, Rossiter-McLaughlin observations have also revealed a significant
population of spin-orbit misaligned planets.  For instance, HAT-P-30b is misaligned
by $74^\circ\pm9^\circ$ \citep{2011ApJ...735...24J} and XO-3 is misaligned by
$70^\circ\pm15^\circ$ \citep{2009IAUS..253..508H} (see \citet{2011exha.book.....P}
Table 6.1 for a summary of spin-orbit alignments).  Many planets are even retrograde,
like HAT-P-11b \citep{2010ApJ...723L.223W,2011PASJ...63S.531H}, WASP-17b
\citep{2010ApJ...709..159A}, and HAT-P-7b
\citep{2009PASJ...61L..35N,2009ApJ...703L..99W}! \citet{2010ApJ...718L.145W} noted
that these highly spin-orbit-misaligned planets occur preferentially around hot
stars, implying that planet formation and/or evolution differs around more massive
stars.

The Rossiter-McLaughlin effect has also been used to determine the spin-orbit
alignment of stellar binaries
\citep{2007A&A...474..565A,2009Natur.461..373A,2011ApJ...726...68A}.  In fortuitous
cases, a star shows spot activity in the area under which the planet transits.  
Those spots' effects on the transit lightcurve can then be used to constrain
spin-orbit alignment \citep{2011ApJ...733..127S,2011arXiv1107.2920S,2011arXiv1107.2106N}.

\citet{2009ApJ...705..683B} showed that the spin-orbit alignment for planets
orbiting rapidly-rotating stars can be determined from transit photometry alone,
taking advantage of the nonuniformity of those stars' disks
\citep{1924MNRAS..84..665V}.  The nonuniformity introduces characteristic
asymmetries into planets' transit lightcurves.  The photometric technique
described by \citet{2009ApJ...705..683B} has an advantage over
Rossiter-McLaughlin measurements in that it can measure both the longitude of the
planet's ascending node and the stellar obliquity to the plane of the sky.  As a
purely radial measurement, Rossiter-McLaughlin determines only $\asc$, the
projected spin-orbit alignment.  The Rossiter-McLaughlin-measured spin-orbit
alignment distribution must be deprojected in order to ascertain the true
distribution \citep{2009ApJ...696.1230F}.  

Recently, \citet{2011ApJ...736L...4S} found an asymmetry in the \emph{Kepler}
transit lightcurve of \planet~that they attributed to spin-orbit misalignment
around a fast-rotating star, consistent with the \citet{2009ApJ...705..683B}
predictions.  In this paper, we use this asymmetry to measure the spin-orbit
misalignment angle $\spinorbit$ of the \planet~system using the
\citet{2009ApJ...705..683B} model.  The present work, focused on the measurement
of $\spinorbit$, leaves to others analysis of the complete lightcurve and
secondary eclipse.  In Section \ref{section:observations} we describe data
reduction of the \emph{Kepler} public Q2 lightcurve for this planet, and we
describe the gravity-darkened rapidly-rotating star transit lightcurve model and
its application in Section \ref{section:model}.  We then show the results of our
fit in Section \ref{section:results} and discuss the implications of the
measurement in Section \ref{section:discussion}.  We give our concluding remarks
in Section \ref{section:conclusion}.

\section{OBSERVATION} \label{section:observations}

The \emph{Kepler} mission \citep{2010Sci...327..977B} science team first
identified a planet candidate around the star with designation 9941662 in the
\emph{Kepler} Input Catalog \citep[KIC,][]{2011arXiv1102.0342B} in the first two
quarters of science data (Q1 and Q2) \citep{2011ApJ...736...19B}.  
A discovery paper confirming the resulting
\emph{Kepler} Object of Interest (KOI) \planet~as a planet is presently in
preparation 
\citep{2011AAS...21710304R}.  
The KIC records the star's location as
19.13141H~$+46.8684^\circ$, and provides stellar parameters of
$M_*=1.83\mathrm{M_\odot}$, $T_\mathrm{eff}=8848\mathrm{~K}$, and a magnitude in
the \emph{Kepler} bandpass of $\mathrm{K_p}=9.958$ \citep{2011ApJ...736...19B}. 
Later additional ground-based spectroscopy by \citet{2011ApJ...736L...4S}
determined $M_*=2.05\mathrm{M_\odot}$, $T_\mathrm{eff}=8511\pm400\mathrm{~K}$,
and a $\Delta$magnitude between \star~and its companion of 0.29. 
\citet{2011ApJ...736L...4S} also measured $v\sin i=65\pm10~\mathrm{km/s}$ for
\star.

We use the \emph{Kepler} photometry, in which apparent transit depth for
\planet~is different in the Q0, Q1, and Q2 \emph{Kepler} time series.  This
interquarter variability (as well as intraquarter variability within Q1) derives
from use of a too-small aperture in Q1 that did not gather all of the flux from
this saturated bright star.  Therefore for our investigation we,
like \citet{2011ApJ...736L...4S}, use just the \emph{Kepler} Q2 public domain
photometry, downloaded from the Multimission Archive at STScI (MAST).  We use
the Q2 short cadence data in our analysis.

In reducing the data first we eliminate data with imperfect trend subtractions
that were acquired near data gaps, leaving 111858 1-minute measurements in the
short cadence time series.  Then we clip out the points whose flux varies by more
than $5\sigma$ away from the 6 surrounding points -- this eliminates 176 bad
measurements.  With no evidence of transit timing variations or other
transit-to-transit variability, we then fold the data using the
\citet{2011ApJ...736...19B} period of 1.763589~days.  We crop the time series
around the transit itself so as not to consider the interesting effects in the
lightcurve other than the transit \citep[\emph{i.e.},][]{AviKOI13}. 
In order to speed computation time in fitting, we bin the data into 60-second
bins, typically coadding 45 points per bin and leaving 254 points in the
timeseries with characteristic precision of 28 ppm.  

Finally, we subtract off 45\% of the median flux in order to compensate for light
from \star's binary companion polluting the photometry 
\citep[following][]{2011ApJ...736L...4S}.

\section{MODEL}\label{section:model}

We generate model lightcurves using the algorithm first described in
\citet{oblateness.2003}, suitably modified for gravity-darkened rotating stars by
\citet{2009ApJ...705..683B}.  It has four different methods to calculate transit
fluxes.  In order of increasing accuracy and required computation time, they
are:  analytical \citep[after][with
gravity-darkening modifications]{2002ApJ...580L.171M}, image-pixel-based
cartesian, radial 1-D numerical
integration (which we call `polar'), and radial/azimuthal 2-D numerical 
integration (which we call `full polar'). 

For this work we also added the following features to the code:

\begin{itemize}

\item \textbf{Fit for $\mathbf{\obliq}$, $\mathbf{\asc}$} -- We added the
ability to fit for the stellar obliquity, $\obliq$, and for the longitude of the
ascending node of the planet's orbit, $\omega$.  In this paper we express
$\omega$ in terms of $\asc$, the sky-projected spin-orbit angle, in order to
provide comparability with existing measurements of other transits from the
Rossiter-McLaughlin effect \citep[\emph{e.g.}][page 128]{2011exha.book.....P}. 
Hence we define $\asc$ to be the longitude of the ascending node as measured
counterclockwise from the ascending node of the star's rotational equator. 
Given the \citet{2011ApJ...736L...4S} measurement of stellar $v\sin i$ (which is
$v\cos\obliq$ using our parameters), we therefore couple the stellar rotation
rate $\Omega_*$ to the stellar obliquity such that:
\begin{equation}\label{eq:omega} \Omega_* = \frac{(v \sin
i)_\mathrm{measured}}{R_* \cos \obliq}~~~~~~. \end{equation}

\item \textbf{Time integration} -- When fitting data, for each model photometric
data point we numerically integrate the flux for $\pm t/2$~around the central
time for the point, where $t$ is that point's integration time.  This has little
effect on the short cadence data, but greatly improves the accuracy of fits
using the 30 minute effective integration time for \emph{Kepler}'s long cadence
data \citep{2010MNRAS.408.1758K}.  

\item \textbf{Bandpass} -- We modified our algorithm to integrate flux numerically
across the published \emph{Kepler} response
function\footnote{http://keplergo.arc.nasa.gov/CalibrationResponse.shtml}. 
Wavelength effects are incorporated entirely into limb darkening parameters for
non-rotating stars.  But fast-rotators' temperature heterogeneity introduces
variations in transit lightcurve shape as a function of wavelength
\citep{2009ApJ...705..683B}.  In practice, for \star, on which we fix the polar
temperature to be 8848K (the value from the KIC), the effects of the bandpass are
nearly indistinguishable from a monochromatic model at $0.5934~\mathrm{\mu m}$.

\item \textbf{Fit for $\mathbf{M_*}$} -- Transit lightcurves for non-rotating stars
can constrain only the stellar density, $\rho_*=\frac{3M_*}{4\pi R_*^3}$.  They
cannot determine  the radius $R_*$ and mass $M_*$ individually because, as can be
derived from Kepler's law, $\frac{M_*}{R_*^3}=\frac{1}{G T_\mathrm{dur}^3 n}$,
where $G$ is the universal gravitational constant, $T_\mathrm{dur}$ is the
transit duration, and $n$ is the planet's orbital mean motion (which is set by
the orbital period).  Since $G$, $T_\mathrm{dur}$, and $n$ are all constant,
there are no independent constraints on $M_*$ or $R_*$ individually.  For a
fast-rotating star, however, the intensity of the gravity darkening effect
depends on the local effective gravity.  For instance, the ratio of the
equatorial stellar effective temperature $T_\mathrm{eq}$ to the polar temperature
$T_\mathrm{pole}$ is
\begin{equation}
\frac{T_\mathrm{eq}^4}{T_\mathrm{pole}^4} =
\frac{R_\mathrm{p}^2}{R_*^2}\big(1-\frac{\Omega^2 R_\mathrm{eq}^3}{GM_*}\big)~~~.
\end{equation}
The $\frac{R_\mathrm{eq}^3}{M_*}$ term provides no new constraints, of course. 
But $\Omega$ depends only on $R_*$, and not on $M_*$ ($\frac{R_\mathrm{p}}{R_*}$
is also a function of $\Omega$ -- see \citet{Seager.oblateness} for how they
relate in a point-core model relevant for stellar moments of inertia).  So
gravity darkening can be used to break the stellar density degeneracy.  

In practice, however, this is a weak constraint.  A stellar mass of
$M_*=0.001~\mathrm{M_\odot}$ can be immediately (if unhelpfully) ruled out for
\star~because even with zero obliquity ($\obliq=0^\circ$) it would need to be
rotating faster than its breakup speed to achieve $v\sin i=65~\mathrm{km/s}$. 
Conversely, a $M_*=100~\mathrm{M_\odot}$ star would need to rotate with a
sufficiently long period $P_{\mathrm{rot}*}=\frac{2\pi}{\Omega}$ that its gravity
darkening would be negligible.  We can derive lightcurve constraints on $M_*$ then
because low-mass stars have exaggerated gravity darkening, and high-mass stars have
muted gravity darkening.  Either can then fail to fit the real lightcurve if the
gravity darkening differs sufficiently from that of the real star.  Highly
asymmetric lightcurves better constrain $M_*$ on the high end, while more nearly
symmetric lightcurves like that for \planet~constrain the low-mass end more
effectively.

\end{itemize}

Lacking empirical constraints, we fix the gravity darkening parameter $\beta$
when we fit the lightcurves.  We use two different assuptions for the value of
$\beta$:  the theoretical value from \citet{1924MNRAS..84..665V} of $\beta=0.25$,
and the experimental value of $\beta=0.19$ from \citet{2007Sci...317..342M}.  As
expected, higher values of $\beta$ lead to more intense gravity darkening holding
all other parameters equal.  The ultimate effect on the transit lightcurve of
changing $\beta$, therefore, is roughly similar to that of varying $M_*$ in that
both primarily affect the intensity of gravity darkening.  We found, however,
that the best-fit values varied less than $1\sigma$ when using $\beta=0.19$
rather than $\beta=0.25$.  We therefore use the theoretical value of $\beta=0.25$
in our reported fits.  More precise photometry and a better determination of the
star's mass could therefore help to constrain $\beta$ empirically in the future.

We fit the model described above to the data using the Leavenberg-Marquardt
algorithm described in \citet{NumericalRecipes} to efficiently zero in on the best
parameters.  Lacking appropriate analytical partial derivatives for most parameters
in the gravity-darkened case \citep[though see also][for analytical derivatives in
the spherical star case]{2008MNRAS.390..281P}, we calculate numerical partial
derivatives with respect to each parameter at each point in each step.  

To calculate the 1-standard deviation ($1\sigma$) errors on each measurement, we
follow a modified version of the set of steps specified by \citet[][page
815]{NumericalRecipes}.  Instead of using the covariance matrix, we explicitly
fit the data using a set of fixed values for each parameter.  We then identified
the minimum in $\chi^2$ as a function of each parameter.  We assign the error
for that parameter based on how far from the minimum value the best-fit $\chi^2$
increases by the appropriate amount given a $1\sigma$ confidence level and our
number of parameters of interest (8).  This $\Delta \chi^2=9.304$ in our case,
since we held $M_*$ fixed when fitting other parameters (see Section
\ref{section:results}).

\section{RESULT}\label{section:results}

\begin{figure}[htb]
\epsscale{1.}
\plotone{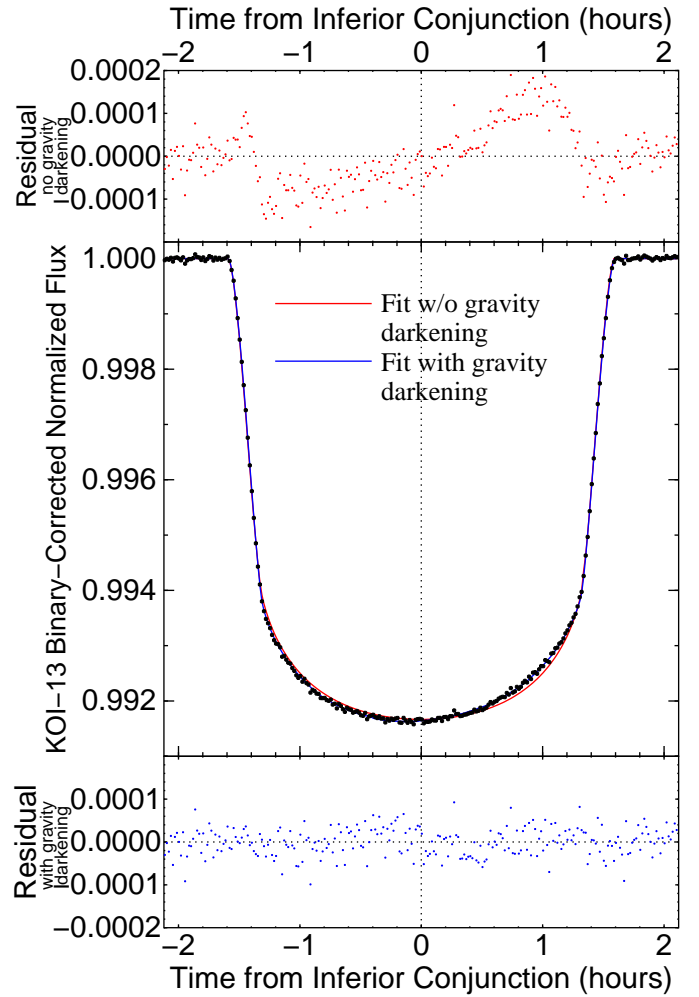}

\caption{\footnotesize This plot shows the transit lightcurve for \planet~in filled black circles
at in the middle, with a best-fit line using gravity darkening
($M_*=2.05~\mathrm{M_\odot}$) in blue and one without gravity darkening in red.  The
size of the data circles corresponds to their error (as reported in the MAST,
adjusted for binning) in the $y$-dimension (flux).   We plot the residuals for the
gravity-darkened fit as blue dots in the bottom graph.  The residuals for the
non-gravity-darkened fit are at top, as red dots. \label{figure:fitgraph}}

\end{figure}

We fit the \emph{Kepler} Q2 short-cadence \planet~transit lightcurve for 9
different parameters:  
\begin{enumerate}
\item the stellar radius, $R_*$;
\item the planetary radius, $R_\mathrm{p}$; 
\item the orbital inclination relative to the plane of the sky, $i$, also expressed as
the impact parameter, $b$;
\item a single limb darkening parameter, $c_1$, equal to the sum of the two
quadratic limb darkening parameters such that $c_1=u_1+u_2$, after
\citet{2001ApJ...552..699B}; 
\item the time of inferior conjunction, $T_0$, which
can be shifted relative to the mid-transit time for oblate stars
\citep{2009ApJ...705..683B};
\item the out-of-transit star flux, $F_0$; 
\item the
stellar mass, $M_*$; 
\item the longitude of the planet's ascending node, $\asc$,
relative to that of the ascending node of the stellar equator; and 
\item the
stellar obliquity, $\obliq$, measured as the tilt of the stellar north
rotational pole toward Earth relative to the plane of the sky.
\end{enumerate}
We show the best-fit values along with their $1\sigma$ uncertainties in Table
\ref{table:bestfit}.  The fit itself along with the data, are plotted in Figure
\ref{figure:fitgraph}.  We also show in Figure \ref{figure:fitgraph} the best-fit
lightcurve without gravity darkening, after \citet{2011ApJ...736L...4S}, in order 
to draw attention to the necessity of the present analysis.  The
no-gravity-darkening fit fails to match the \planet~transit ingress and egress.  It
also does not adequately model the transit bottom, which for \planet~is darker in
the first half of the transit than it is in the second half.  The gravity-darkened
model fits all of these characteristics -- the ingress, egress, and
asymmetric transit bottom.

\begin{center}
\begin{table}[!htbp]
\begin{tabular}{|r|c|c|}
\hline
          & \multicolumn{2}{|c|}{Best-Fit Values}  \\
Parameter & $M_*=1.83~\mathrm{M_\odot}$           & $M_*=2.05~\mathrm{M_\odot}$  \\
\hline
$\chi^2_\mathrm{reduced}$ & 1.409 & 1.419 \\ \hline
$R_*$ & $1.694\pm0.013~\mathrm{R_\odot}$ & $1.756\pm0.014~\mathrm{R_\odot}$\\ \hline
$R_\mathrm{p}$ & $1.393\pm0.015~\mathrm{R_{Jup}}$ & $1.445\pm0.016~\mathrm{R_{Jup}}$  \\ \hline
$\frac{R_\mathrm{p}}{R_*}$ & 0.084508 & 0.084513 \\ \hline
$i$ & $85.9^\circ\pm0.4^\circ$ &  $85.9^\circ\pm0.4^\circ$  \\ \hline
$b$ & 0.31962 & 0.31598 \\ \hline
$c_1$ & $0.49\pm0.03$ & $0.48\pm0.03$ \\ \hline
$T_0$ & $4628032\pm3$~s & $4628033\pm3$~s \\ \hline
$\asc$ & $24^\circ\pm4^\circ$ & $23^\circ\pm4^\circ$ \\ \hline
$\obliq$ & $-45^\circ\pm4^\circ$ &  $-48^\circ\pm4^\circ$ \\ \hline
$\spinorbit$ & $54^\circ\pm4^\circ$  & $56^\circ\pm4^\circ$ \\ \hline
$P_{\mathrm{rot}*}$ & 22.5~hr& 22.0~hr\\ \hline
$f_*$ & 0.018 & 0.021 \\ 
\hline
\end{tabular}
\caption{\footnotesize \label{table:bestfit}
Best-fit transit parameters for the \planet~system.
The time of inferior conjunction, $T_0$, is measured in seconds after
BJD~2454900, after \citet{2011ApJ...736...19B}.  There is a 4-way degeneracy for the
transit geometry, \emph{i.e.} the projected spin-orbit angle $\asc$ and the stellar 
obliquity $\spinorbit$ -- see Figure \ref{figure:geometry}.  The stellar rotation
period is denoted $P_{\mathrm{rot}*}$.  The stellar dynamical oblateness is denoted $f_*$.
}

\end{table}
\end{center}

The best-fit value for stellar mass $M_*$ is highly uncertain due to the weak
dependence of gravity darkening on $M_*$ as discussed in Section
\ref{section:model}.  Both the \citet{2011ApJ...736...19B} and
\citet{2011ApJ...736L...4S} spectroscopic measurements of $M_*$,
$M_*=1.83~\mathrm{M_\odot}$ and $M_*=2.05~\mathrm{M_\odot}$ respectively, lie
within $2\sigma$ of our measured value of $M_*=0.9\pm0.6~\mathrm{M_\odot}$.  Our
measurement clearly prefers lower values for $M_*$.  However, the best-fit value
of $M_*=0.9~\mathrm{M_\odot}$ is clearly inconsistent with the star's spectral
type.  Given the uncertainty in our measurement, however, for the remainder of
the fitted parameters we show values for an assumed mass of both
$M_*=1.83~\mathrm{M_\odot}$ \citep{2011ApJ...736...19B} and
$M_*=2.05~\mathrm{M_\odot}$ \citep{2011ApJ...736L...4S}.

The reduced $\chi^2$ ($\chi^2_\mathrm{reduced}$) values for each of those fits
are $\chi^2_\mathrm{reduced}=1.409$ ($M_*=1.83~\mathrm{M_\odot}$) and
$\chi^2_\mathrm{reduced}=1.419$ ($M_*=2.05~\mathrm{M_\odot}$).  We think that
the $\chi^2_\mathrm{reduced}$ above 1.0 results from inter-transit variability
in the \emph{Kepler} photometer.  Although each individual data point from the
\emph{Kepler} Q2 dataset has high precision, red noise on longer timescales
leads to systematic errors between transits.  Hence folding the transits
introduced a modest amount of additional noise over and above the precision of
the individual points.

We adjusted the formal $1\sigma$ error bars to account for the final
$\chi^2_\mathrm{reduced}$ of each fit.  The errors that we indicate in Table
\ref{table:bestfit} represent the formal errors of the fit.  They do not include,
for instance, likely systematic errors from our assumptions of stellar mass,
$T_\mathrm{pole}$, the binary flux fraction, and/or fixed $v\sin i$.

Our measured value for the stellar radius that we derive from the lightcurve fit,
$R_*=1.756\pm0.014~\mathrm{R_\odot}$, is substantially lower than the spectroscopic
value reported by \citet{2011ApJ...736L...4S}.  Given a similar planet-star
radius ratio, then, we also measure a smaller radius for the transiting body,
$R_\mathrm{p}=1.445\pm0.016~\mathrm{R_{Jup}}$.  This smaller radius for the
transiter places it within the range of possible planetary companions for relatively
young, highly-irradiated planets \citep{2007ApJ...659.1661F}.  According to Figure 5
from \citet{2007ApJ...659.1661F}, both $M_\mathrm{p}=1~\mathrm{M_J}$ and
$M_\mathrm{p}=3~\mathrm{M_J}$ planets with $25~\mathrm{M_\oplus}$ cores could have
radii of $R_\mathrm{p}=1.44~\mathrm{R_{Jup}}$ with an age of 700 Myr as determined
by \citet{2011ApJ...736L...4S}.  This raises the possibility that \planet's mass
lies in the planetary regime, not that of brown dwarfs as the higher $R_\mathrm{p}$
of \citet{2011ApJ...736L...4S} suggested.

By fitting for both $R_*$ and our limb darkening parameter $c_1$, we arrive at a
more consistent measure of the planet's orbit inclination relative to the plane of
the sky, $i$.  Expressed as an impact parameter $b$, our values near $b=0.32$ are
significantly lower (\emph{i.e.} the transit is more nearly central, and less close
to grazing) than the value determined by \citet{2011ApJ...736L...4S}.  The
discrepancy may lie in the assumption by \citet{2011ApJ...736L...4S} of a
spectroscopically-determined stellar radius that was rather higher than our best-fit
value.  Because the stellar oblateness, $f_*$, affects ingress and egress times
(similar to the way that planetary oblateness does, see \citet{oblateness.2003}),
our measured impact parameter is more accurate than one that would result from a fit
without explicitly incorporating stellar oblateness.

Stellar oblateness $f_*$ also affects the measured time of inferior conjunction,
$T_0$.  For rapidly-rotating, oblate stars, the mid-transit time and the time of
inferior conjunction can differ \citep{2009ApJ...705..683B}, depending on the
transit geometry.  The measured times for the fit assuming
$M_*=1.83~\mathrm{M_\odot}$ ($4628032\pm3$~s after BJD~2454900) and that while
assuming $M_*=2.05~\mathrm{M_\odot}$ ($4628033\pm3$~s after BJD~2454900) are
slightly different because the intensity of the effect depends on the stellar
oblateness, which is different in the two fits (see below).  Both values are
somewhat later than the mid-transit time reported in
\citet{2011ApJ...736...19B}, $4628014\pm10~\mathrm{s}$.  Since the values are within
$2\sigma$, however, the discrepancy could be due to the stellar oblateness, but may
also be statistical variation.

The transit geometry determinations represent the most important measurements
that we describe in this paper.  Our value for the sky-projected spin-orbit
angle $\asc$ can be directly compared to measurements of $\asc$ made in other
transiting systems from the Rossiter-McLaughlin effect. Unlike
Rossiter-McLaughlin measurements, however, we are also able to determine the
stellar obliquity, $\obliq$, which we define as the amount by which the star's
north pole is tipped out of the plane of the sky toward the observer.

\begin{figure}[htb]
\epsscale{1.}
\plotone{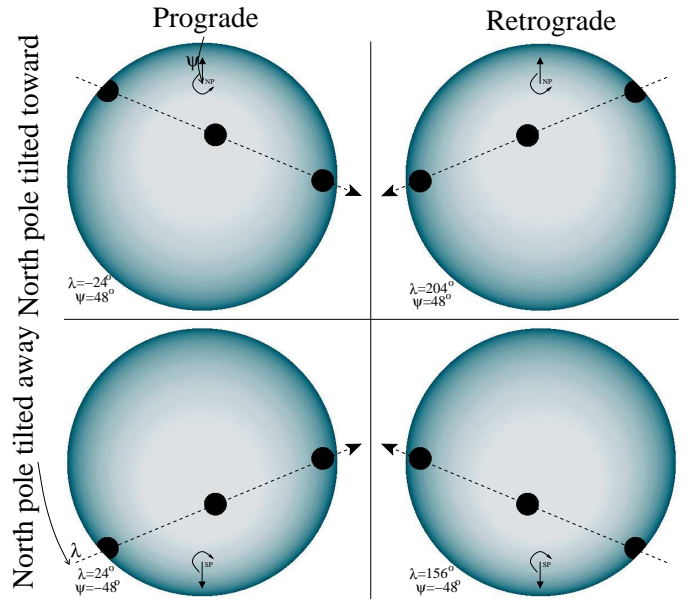}
\caption{\footnotesize Allowed geometries for the \planet~transit event.  Best-fit values for an
assumed $M_*=2.05~\mathrm{M_\odot}$ are shown.  We enhance the contrast of the
stellar image in order to make the heterogeneity of the disk more evidently visible.
Note that due to the competing effects of limb darkening and gravity darkening, the
stellar poles are not the brightest part of the stellar disk.  Gravity darkening
does, however, displace the brightest part of the disk poleward from the center of
the disk.  ``NP" and ``SP" stand for North Pole and South Pole, respectively.
\label{figure:geometry}}
\end{figure}

We cannot determine $\asc$ and $\obliq$ uniquely, though.  We determine 4
allowed transit geometries, as diagrammed in Figure \ref{figure:geometry}.  For
the $M_*=2.05~\mathrm{M_\odot}$ case, those are (clockwise from lower-left): 
(1) $\asc=24^\circ$, $\obliq=-48^\circ$, the prograde case with north pole
tipped away from the observer; (2) $\asc=-24^\circ$, $\obliq=48^\circ$, the
prograde case with the north pole tipped toward the observer; (3)
$\asc=-156^\circ$, $\obliq=48^\circ$, the retrograde case with the north pole
tipped toward the observer; and (4) $\asc=156^\circ$, $\obliq=-48^\circ$, the
retrograde case with the north pole tipped away from the observer.

The combination of $\asc$ and $\obliq$, along with the planet's orbit inclination
$i$, allow us to calculate the complete relative angle between the stellar rotation
pole and the planetary orbit normal -- the spin-orbit alignment, $\spinorbit$ --
using \citet{2007AJ....133.1828W} Equation 7.  With our angle definitions, this
equation becomes
\begin{equation}
\cos\spinorbit = \sin\obliq\cos i + \cos\obliq\sin i \cos\asc~~~~~.
\end{equation}
We measure $\spinorbit=54^\circ\pm4^\circ$ for $M_*=1.83~\mathrm{M_\odot}$, and
$\spinorbit=56^\circ\pm4^\circ$ for $M_*=2.05~\mathrm{M_\odot}$.  Hence the orbiting
body, \planet, is far from spin-orbit alignment.  While the projected alignment
$\asc$ (as would be measured using the Rossiter-McLaughlin effect) alone reveals a 
significant deviation from alignment, the full extent of the misalignment manifests
from the large stellar obliquity $\obliq$ which would not be measured by the
Rossiter-McLaughlin effect \citep[\emph{i.e.},][]{2009ApJ...696.1230F}.

In actuality, given the orbital geometry degeneracies described by Figure
\ref{figure:geometry}, $\spinorbit$ also has another solution -- one where the
planet orbits retrograde.  In that case, $\spinorbit=126^\circ\pm4^\circ$ for
$M_*=1.83~\mathrm{M_\odot}$, and $\spinorbit=124^\circ\pm4^\circ$ for
$M_*=2.05~\mathrm{M_\odot}$.  We offer a possibility for resolving the
prograde/retrograde degeneracy photometrically in Section \ref{section:discussion}.

\section{DISCUSSION}\label{section:discussion}

\subsection{Implications}

Our determination of a high spin-orbit misalignment for \planet, $\sim55^\circ$
rather than the solar system's more typical $\leq6^\circ$, drives us to pose the
question of how the spin-orbit misalignment came to be.  Assuming that
\planet~formed like a solar system planet, in a planar disk orbiting the
protostellar \star, there are two ways to have developed a spin-orbit misalignment: 
either the stellar spin axis changed, or the planet's orbit did (or both).  In
addition to explaining the the spin-orbit misalignment of \planet, any explanation
for its formation and evolution must also explain \planet's apparently circular
orbit, as established by the presence of a properly-timed secondary eclipse
\citep{2011ApJ...736L...4S}.  

\citet{2010MNRAS.401.1505B} suggest a mechanism by which to alter the star's
rotation axis, moving it away from the normal to the protoplanetary disk.  This
mechanism could drive spin-orbit misalignment, and allows for \planet~to end up on a
circular orbit.

Two prominent ways to alter the planet's orbit are by planet-planet scattering and
the Kozai mechanism.  If a planet-planet scattering event occurred early enough in
the planet formation process, then sufficient debris might have remained to
circularize the orbit before the disk dissipated.  

A Kozai-driven source for spin-orbit misalignment, like that for HD80606b
\citep{2003ApJ...589..605W,2009ApJ...703.2091W}, is appealing because of the
presence of \star's binary companion.  However this mechanism requires that the
planet's orbital precession due to the binary companion be more rapid than that
induced by the planet's parent star alone.  \citet{2003ApJ...589..605W} suggest
that HD80606b's Kozai variations thus ended when its orbital semimajor axis
dropped, reducing Kozai's effectiveness while increasing the effects from its
parent star. 

For this mechanism to be viable for the evolution of \planet, its orbit would
have had to have been circularized from a presumed prior highly-eccentric
Kozai-derived orbit.  Given \star's young 700~Myr age, though, tidal
circularization would seem difficult.  The conventional tidal circularization
timescale $\tau_\mathrm{c}=\frac{e}{\mathrm{d}e/\mathrm{d}t}$
\citep{2000ApJ...537L..61T} for \planet~in its present state is
$\tau_\mathrm{c}=\sim8~\mathrm{Myr}$.  However in a presumed former state like
that of HD80606b presently, with a semimajor axis of $0.5~\mathrm{AU}$, the
circularization timescale is $\tau_\mathrm{c}=200$ trillion years using the value
for $M_\mathrm{p}$ from \citet{AviKOI13}.  Hence we think that the Kozai
mechanism is likely not responsible for the spin-orbit misalignment of \planet,
since the end state of Kozai migration would be a highly eccentric orbit that
could not have been circularized in the star's lifetime.  A more complete
numerical study of the orbital circularization of \planet, tracking the coupled
evolution of both eccentricity and semimajor axis following
\citet{2008ApJ...678.1396J}, might shed additional light on the issue.

\subsection{Future Work}

The present work used just the public Q2 short-cadence \emph{Kepler} data for
\star.  Presuming that \emph{Kepler} continues to observe \star~at short cadence,
the ultimate coadded photometric precision should continue to improve.  The
resulting higher signal-to-noise ratio will allow more precise determinations of all
of the transit parameters in Table \ref{table:bestfit}.  As the precise shape of the
asymmetry at the bottom of the lightcurve becomes clearer, the purely photometric
determination of the star's mass $M_*$ should approve commensurately.

Another effect that will become important as precision improves is the
Photometric Rossiter-McLaughlin effect
\citep{2011arXiv1107.4458S,2011arXiv1104.3428G} -- the photometric analog of the
Rossiter-McLaughlin effect.  This effect arises due to the rotation of the star
combined with the beaming effect.   During transit the planet covers up varying
parts of  the star, with different apparent radial velocities.  The different
radial velocities cause photons emitted from different parts of the stellar
surface to be beamed by a different amount, resulting in an anomalous
photometric signal during transit.  This combined effect has been proposed to
investigate the spin-orbit alignment of \emph{Kepler}'s binary star population
\citep{2011arXiv1107.4458S}.

\begin{figure}[htb]
\epsscale{1.}
\plotone{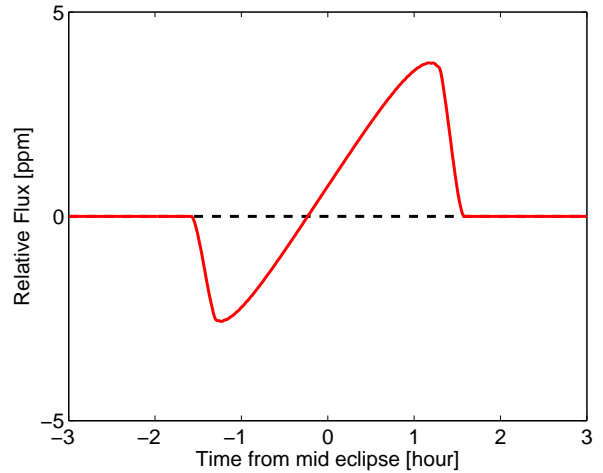}
\caption{\footnotesize This plot shows the intensity of the Photometric Rossiter-McLaughlin effect
for \planet~using our best-fit parameters for $M_*=2.05~\mathrm{M_\odot}$ from Table
\ref{table:bestfit}.  While our algorithm presently does not account for this
effect, it could be incorporated in the future as additional
\emph{Kepler} data improves the period-folded photometric precision.  If it could be detected, the Photometric
Rossiter-McLaughlin effect could be used to break the prograde/retrograde degeneracy
for the planet's orbit depicted in Figure \ref{figure:geometry}.
\label{figure:photometricRM}}
\end{figure}

We show a calculation of the intensity of this effect for \planet~in Figure
\ref{figure:photometricRM}.  While its present amplitude of $\sim4$~ppm renders
it  undetectable in the present Q2 \emph{Kepler} data, future improvements in
precision as the mission progresses will make the effect detectable and more
easily discernible from the gravity-darkening effects addressed in this paper. 
Most importantly, the sign of the Photometric Rossiter-McLaughlin effect would
be flipped for retrograde-orbiting planets.  Hence the Photometric
Rossiter-McLaughlin effect holds promise in the future for a purely photometric
resolution to the prograde/retrograde degeneracy that we show in Figure
\ref{figure:geometry}.

\begin{figure}[htb]
\epsscale{1.}
\plotone{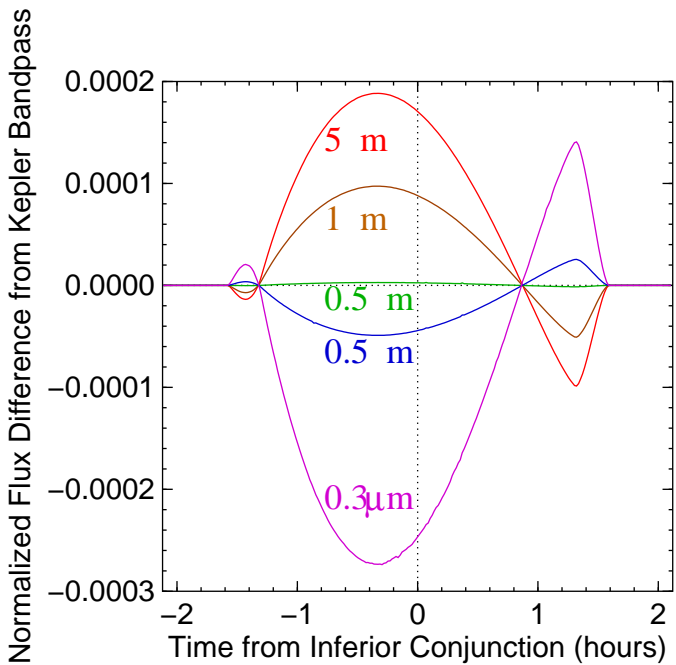}
\caption{\footnotesize Here we show the difference between various theoretical lightcurves,
conforming to the transit parameters from Table \ref{table:bestfit}, acquired at
various indicated monochromatic wavelengths relative to that integrated across the
\emph{Kepler} bandpass.  At shorter wavelengths the \planet~transit lightcurve
asymmetry is increased, and at longer wavelengths the asymmetry is more muted.  A single-wavelength lightcurve at $0.5934~\mathrm{\mu m}$ 
most closely replicates that of the lightcurve under the entire 
\emph{Kepler} bandpass for \star.  Observation of this transit, if sufficiently
precise, could constrain $T_\mathrm{pole}$ and the gravity-darkening parameter
$\beta$.
\label{figure:colorgraph}}
\end{figure}

Finally, \citet{2009ApJ...705..683B} showed that the lightcurve shape varies for
spin-orbit misaligned planets when viewed at differing wavelengths.  We show an
illustration of this effect for \planet~in Figure \ref{figure:colorgraph}. 
Because the intensity of the difference between the lightcurve in the
\emph{Kepler} bandpass and that at different wavelengths depends on
$T_\mathrm{pole}$, such observations could serve to place constraints on
$T_\mathrm{pole}$.  

\section{CONCLUSION}\label{section:conclusion}

\citet{2009ApJ...705..683B} first discussed the prospect for spin-orbit misaligned
planets orbiting rapidly-rotating stars having asymmetric transit lightcurves owing
to stellar gravity darkening \citep{1924MNRAS..84..665V}. 
\citet{2011ApJ...736L...4S} showed that planet candidate \planet, first identified
in \emph{Kepler} data by \citet{2011ApJ...736...19B}, has an asymmetric transit
lightcurve.  \citet{2011ApJ...736L...4S} also made new critical measurements of
\star, including the binary flux fraction and $v\sin i$ from stellar rotation.

Fitting the Q2 short cadence \emph{Kepler} photometry for \planet~with the
\citet{2009ApJ...705..683B} model, we measured a projected spin-orbit alignment
$\asc=23^\circ\pm4^\circ$ and a stellar obliquity (defined as the angle by which the
stellar north pole is tilted out of the plane of the sky toward the observer) of
$\obliq=-48^\circ\pm4^\circ$.  While these measurements assume
$M_*=2.05~\mathrm{M_\odot}$ as determined by \citet{2011ApJ...736L...4S}, other fits
using different assumed stellar masses show that the overall spin-orbit misalignment
derived for the star, $\spinorbit=56^\circ\pm4^\circ$, is not substantially affected
by plausible stellar mass variations.  The photometric determination also allows for
a retrograde orbit with $\spinorbit=124^\circ\pm4^\circ$.  Our measurement of
$\spinorbit$ is the first spin-orbit determination to come from gravity darkening.
It is one of just a handful of
observations of any kind to determine the complete spin-orbit misalignment of a
transiting planetary-class body, not just its projected spin-orbit alignment $\asc$
\citep{2007AJ....133.1828W,2011arXiv1107.2920S,2011arXiv1107.2106N}.

The spin-orbit misalignment of \planet~follows the trend that planets around more
massive stars are more likely to be spin-orbit misaligned
\citep{2010ApJ...718L.145W}.  Our \planet~measurement together with spin-orbit
misaligned WASP-33b around a $M_*=1.5~\mathrm{M_\odot}$ star
\citep{2010MNRAS.407..507C} extends the  \citet{2010ApJ...718L.145W} to yet-higher
stellar masses.

We also measure the other critical transit parameters for the \planet~system.  We
determined $R_*=1.756\pm0.014~\mathrm{R_\odot}$,
$R_\mathrm{p}=1.445\pm0.016~\mathrm{R_{Jup}}$, $i=85.9^\circ\pm0.4^\circ$,
$b=0.31598$, and a single limb darkening parameter of $c_1=0.48\pm0.03$.  Our
photometrically-determined stellar and planetary radii are much lower than those of
\citet{2011ApJ...736L...4S} who used a spectroscopically-derived star radius of
$2.44~\mathrm{R_\odot}$.  Our lower estimated radius places \planet~in the regime
where interior models allow for highly-irradiated planetary-mass objects
\citep{2007ApJ...659.1661F}.

We also take advantage of the dependence of gravity darkening on the stellar
surface gravity to directly measure $M_*$ from the \emph{Kepler} transit
lightcurve.  Transit lightcurves for spherical stars can only constrain the
stellar density $\rho_*$, and do not provide any absolute length scale with which
to constrain $M_*$.  Smaller stars would need to rotate faster to provide the
measured $v\sin i=65~\mathrm{km/s}$, and thus they drive more intense curvature
at the bottom of the lightcurves of transiting objects.  Conversely, more massive
stars have correspondingly lower gravity darkening, and less lightcurve
asymmetry.  The effect is weak enough that our highly uncertain measurement of
$M_*=0.9\pm0.6~\mathrm{M_\odot}$ cannot differentiate between the
\citet{2011ApJ...736...19B} and \citet{2011ApJ...736L...4S} spectroscopic
measurements of $M_*=1.83~\mathrm{M_\odot}$ and $M_*=2.05~\mathrm{M_\odot}$
respectively.  While it is consistent with (or within $2\sigma$ of) either
measurement, and inconsistent with the measured spectral type (A), we do weakly
prefer lower values of $M_*$.  Future observations, if more precise, can help to
more tightly constrain this first-ever measurement of $M_*$ from gravity
darkening.

Additional \emph{Kepler} photometry may also show evidence for the Photometric
Rossiter-McLaughlin effect \citep{2011arXiv1107.4458S,2011arXiv1104.3428G} in
the \planet~transit lightcurve.  This sign of this effect, if it can be
measured, would resolve the degeneracy between prograde and retrograde planetary
orbits using photometry alone.

We have not determined the cause of \planet's spin-orbit misalignment.  However we
note that the Kozai mechanism, implicated for the evolution of HD80606b
\citep{2003ApJ...589..605W}, may be inconsistent with \planet's circular orbit 
according to a simple analysis, despite
the presence of a similar-mass binary companion to \star.  A more sophisticated 
study of the effectivity of the Kozai mechanism for \planet~can help to resolve 
the origin of the spin-orbit misalignment.  Stellar spin drift 
and planet-planet scattering during formation remain viable misalignment generation
mechanisms.

The complex and interesting nature of the \planet~star-planet system make it an
astrophysical laboratory for photometric transit effects.  The exquisite photometric
precision of the \emph{Kepler} photometer allows us to probe those effects for the
first time.  The lessons that we learn here will then be applicable to future
transiting objects discovered by \emph{Kepler} and others.

\acknowledgements

\bibliographystyle{apj}
\bibliography{references}

\end{document}